\documentclass[twocolumn,trackchanges]{aastex62}

\graphicspath{{./}{figures/}}

\received{XXXX}
\revised{XXXX}
\accepted{XXXX}

\shorttitle{Star formation in IC 2574}
\shortauthors{Mondal et al.}

\begin{document}

\title{UVIT view of dwarf irregular galaxy IC 2574 : Is the star formation triggered due to expanding H$~$I shells?}

\author{Chayan Mondal}
\affiliation{Indian Institute of Astrophysics, Koramangala II Block, Bangalore-560034, India}
\affiliation{Pondicherry University, R.V. Nagar, Kalapet, 605014, Puducherry, India}

\email{chayan@iiap.res.in, mondalchayan1991@gmail.com}

\author{Annapurni Subramaniam}
\affiliation{Indian Institute of Astrophysics, Koramangala II Block, Bangalore-560034, India}

\author{Koshy George}
\affiliation{Department of Physics, Christ University, Bangalore, India}

\keywords{galaxies: dwarf – galaxies: individual – galaxies: star formation}

\begin{abstract}
Star forming dwarf galaxies in the local volume are diverse and are ideal test beds to understand details of star formation in a variety of environments. Here, we present a deep FUV imaging study of a nearby dwarf irregular galaxy IC 2574 using the Ultraviolet Imaging Telescope (UVIT). We identified 419 FUV bright regions with radii between 15 - 285 pc in the galaxy and found that 28.6\% of them to be located in H~I shells, 12.6\% inside holes and 60.1\% to be away from the holes. The H~I column density is found to be more than $10^{21} cm^{-2}$ for 82.3\% of the identified regions. 30 out of the 48 H~I holes show triggered star formation in their shells while 16 holes do not show any related FUV emission. Cross-matching with H$\alpha$ emission, we found that 23 holes have both FUV and H$\alpha$ emission in their shells signifying very recent trigger. Therefore, star formation in the galaxy has been partly triggered due to the expanding H~I holes whereas in majority of the sites it is driven by other mechanisms. Irrespective of the location, larger star forming complexes were found to have multiple sub-structures. We report two resolved components for the remnant cluster of the super giant shell and estimated their masses. The star formation rate of IC 2574 is found to be 0.57 $M_{\odot}$/yr, which is slightly higher compared to the average value of other nearby dwarf irregular galaxies.
\end{abstract}

\section{Introduction}
Star forming dwarf galaxies constitute one of the very common type of galaxies present in the local volume and increase in number and importance with redshift. They mostly have small intrinsic size and low absolute luminosity \citep{hodge1971}. Study of nearby dwarf galaxies are vital to throw light on the nature of star formation in low mass and metal poor environment \citep{weisz2011}. Dwarf galaxies show a wide variety of star formation history in terms of star formation rate (SFR) and the most recent peak in their star forming activity \citep{cignoni2018}. Perturbations due to stellar feedback, ram pressure stripping or tidal interaction can significantly affect the evolution of a dwarf galaxy \citep{revaz2018}. Many dwarfs are known to undergo bursts of star formation, with the duration of star bursts  extending up to 1000 Myr in some cases. These local star bursts play a deciding role in the evolution of the galaxy in different ways. Ionizing radiation, stellar wind and supernova explosion resulting from massive stars formed in the burst, can disrupt the available gas and also alter the chemical composition of the host galaxy \citep{mcquinn2010a,mcquinn2010b}. There are different mechanisms which can trigger a local burst of star formation in a dwarf galaxy. Expanding H$~$I shell can act as a trigger for secondary star formation in a galaxy \citep{tagle2005}.\\

The H$~$I shells are shell like structure created in the interstellar medium (ISM) of a galaxy due to the combined effect of stellar wind and supernova explosion \citep{weaver1977,mccray1987,tenorio1988,chu2004,bagetakos2011}. There are also other theories like gamma ray burst \citep{loeb1998} or infall of high velocity clouds \citep{tenorio1988,murray2004}, which are proposed as the cause behind the creation of such shells. Recent study by \citet{weisz2009} shows that multiple star formation events can also provide the required energy to form shells in the ISM. H$~$I shells were first identified in the Milky Way by \citet{heiles1979}. The diameter of these shells can vary from 10 pc to more than 1000 pc. A shell with diameter more than 1 kpc are usually called as super giant shell (SGS) \citep{stewart2000}. SGSs are mostly found in dwarf galaxies, where they can sustain for a longer time due to their slow solid body-like rotation and the absence of spiral density wave. Due to the low ambient density and shallow potential well in dwarfs, a shell can also expand to larger radii. Understanding such shells is importantly relevant in the context of recent star formation in the host galaxy. The evolution of shells and their impact in the ISM of the host galaxy in terms of feedback and metal enrichment are equally important research topics. Expanding H$~$I shells have significant effect on to the ISM of a galaxy in local scales. The feedback due to the shell expansion can sweep up the ambient ISM and compress it to trigger further star formation. \citet{kawata2014} reported that star formation can also be triggered due to the collision of such expanding shells in dwarf galaxies. \citet{hopkins2008} showed that the accumulation of neutral and molecular gas along the wall of such shells can trigger further star formation. Studies on nearby dwarf galaxies, such as the Large Magellanic Cloud (LMC), IC 10, Holmberg I, Holmberg II, have confirmed the presence of star formation triggered due to expansion or collision of H~I shells \citep{yamaguchi2001,leroy2006,egorov2017,egorov2018}.\\

IC 2574 (also known as DDO 81 or UGC 5666) is a gas-rich dwarf irregular galaxy in the M81 group, located at a distance of 3.79 Mpc \citep{dalcanton2009}. The galaxy, being present $\sim$ 164 kpc away from the brightest group member M81, does not have any signature of interaction \citep{yun1999}. Parameters of this galaxy are listed in Table \ref{ic2574}. The SFR of the galaxy is found to have increased during the last 1 Gyr \citep{walter1999} and hence it is a good candidate to study recent enhancement in the star forming activities. The ISM of this galaxy is very interesting and intriguing. The distribution of H$~$I in this galaxy shows many shell like structures. Using Very Large Array (VLA) observations, \citet{walter1999} identified 48 H$~$I shells and holes in the galaxy and studied them in detail. In order to understand the connection between H~I holes and triggered star formation in the galaxy, they used the catalog of H$\alpha$ emitting regions from \citet{miller1994} and defined 40 H~II regions, and investigated their connection with the features observed in the H~I map. They reported that the H$\alpha$ emissions, which traces the current star formation, are predominantly found along the rims of larger holes. \citet{pasquali2008} used optical images taken with Large Binocular Telescope (LBT) to explore the star formation history of the galaxy and reported two recent bursts of star formation in the galaxy. They found an older burst about 100 Myr ago inside 4 kpc radius and another younger burst during last 10 Myr between radius 4 to 8 kpc. They further noticed that the younger burst of star formation is mainly located in the periphery of H$~$I shells. Therefore the galaxy IC 2574 offers a good opportunity for studying nature of star formation triggered due to the expanding H$~$I shells. \citet{pellerin2012} observed the north-eastern part of the galaxy using Hubble Space Telescope (HST) and detected 75 young stellar groups (age $\sim$ 10 Myr), which again points to the recent star formation in the galaxy.\\

Young star forming regions in galaxies often produce OB associations, a loosely bound group of O and B type stars \citep{melnik1995}. These young and massive OB stars produce copious amount of far-UV (FUV) radiation, which is an excellent proxy to trace regions of recent star formation activity and to explore the triggering mechanisms. One of the identified SGSs, located in the north-east of the galaxy, has been studied rigorously in the literature. Using UV observations from UIT, \citet{stewart2000} identified a remnant cluster inside the SGS and confirmed a causal relation between the remnant cluster and the triggered star formation seen along the rim of the shell. \citet{cannon2005} used Spitzer observations to conclude that the triggered star formation detected along the rim of the shell is found to alter the temperature of the dust present around it. An HST study by \citet{weisz2009} also revealed that star formation happened earlier inside the SGS has triggered further recent star formation along its rim. \citet{yukita2012} identified a luminous X-ray source, which is slightly offset from the remnant cluster, inside the SGS. They also showed GALEX FUV image of the SGS (their Figure 2) where the remnant cluster is shown as C2. Though the SGS located in the north-eastern part of the galaxy is studied in FUV, a detailed investigation of the FUV emission with respect to all the H~I holes in IC 2574 has not been performed.\\    

In this study, a deep FUV image of this galaxy, obtained from the Ultra-Violet Imaging Telescope (UVIT), is used to understand the connection between the recent star formation and the H~I holes across the galaxy. \citet{walter1999} did a similar study connecting the  H$\alpha$ emission and H~I holes, whereas we combine the FUV emission along with the H$\alpha$ emission to trace the star formation not only in and around  H~I holes, but also in the entire galaxy. Here we trace star forming clumps in the FUV and check for their possible connection with H~I holes, whereas the optical study of \citet{pasquali2008} investigated star formation across the galaxy and its overall connection with the H~I holes. Our study thus aims to present a comprehensive view of the recent star formation in this galaxy, particularly using the FUV images. UVIT on-board  AstroSat \citep{kumar2012} is capable of imaging in UV with a better spatial resolution than GALEX. \citet{mondal2018,koshy2018a,koshy2018b,koshy2018c} used UVIT observations to understand different characteristic of star formation in some selected galaxies and demonstrated the capability of UVIT in utilising better spatial resolution. We make use of the resolution of UVIT ($\sim 0.4\arcsec/$pixel) in the FUV to identify the star forming regions. We identify complexes of possible OB associations and their location with respect to the H~I holes and shells. We study their correlation with the H~I density as well as structural properties, and estimate size, FUV flux and SFR density. The paper is arranged as follows. The observations and data are presented in Section \ref{data_s}, extinction in UV in Section \ref{ext_s}, analysis in Section \ref{analysis_s} followed by discussions and summary in Section \ref{discussion_s} and \ref{sumarry_s} respectively.

\begin{table}
\centering
\caption{Properties of IC 2574}
 \label{ic2574}
\resizebox{90mm}{!}{
\begin{tabular}{ccc}
\hline
 Property & Value & Reference\\\hline
 RA & 10 28 23.5 & \citet{skrutskie2006}\\
 DEC & +68 24 43.7 & \citet{skrutskie2006}\\
 Distance & 3.79 Mpc & \citet{dalcanton2009}\\
 Metallicity (Z) & $0.006$ & \citet{cannon2005}\\
 Inclination & $63^\circ$ & \cite{pasquali2008}\\
 PA of major axis & $55^\circ$ & \citet{pasquali2008}\\
 H~I mass & $14.8\times10^8$ $M_{\odot}$ & \citet{walter2008}\\\hline
\end{tabular}
}
\end{table}

\section{Observations and Data} 
\label{data_s}
\begin{figure*}
\begin{center}
\includegraphics[width=7in]{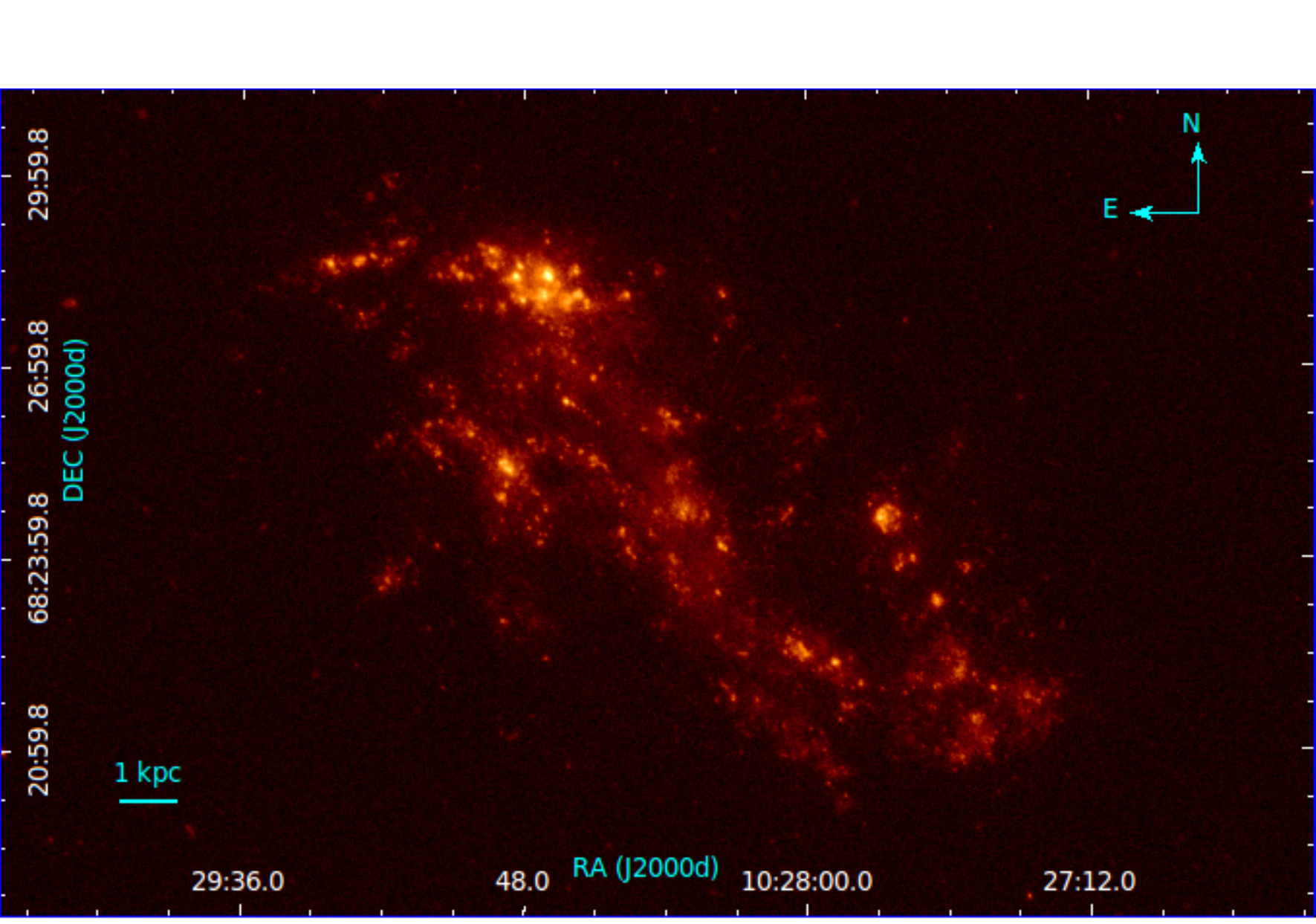} 
 \caption{The FUV image of the galaxy IC 2574 observed with UVIT F148W filter.}
 \label{uv_disk}
 \end{center}
 \end{figure*}

We performed a deep FUV imaging observation of the galaxy IC 2574 using the UVIT on-board AstroSat satellite \citep{kumar2012}. The instrument consists of two co-aligned telescopes, one for FUV (1300 - 1800 $\AA{}$) and another for both NUV (2000 - 3000 $\AA{}$) and Visible. It has the capability of observing sources simultaneously in all the three available bands. The visible channel is mainly used to track the drift pattern in the image produced due to the motion of the satellite. Each of the telescopes has a 28 $\arcmin$ circular field of view with an angular resolution of 1.4 $\arcsec$ in FUV and 1.2 $\arcsec$ in NUV. Both the FUV and NUV channels of UVIT are equipped with multiple photometric filters which provide a unique imaging capability in the ultra-violet bands. The galaxy IC 2574 was imaged in F148W, an FUV broad band filter of UVIT. The data in the NUV channel was not made available due to payload related problems. The observations were carried out with 9 orbits of AstroSat. The images of each orbit are drift corrected, aligned and combined with the help of a customized software CCDLAB \citep{postma2017} to produce a final deep image of 10375 seconds exposure time (Figure \ref{uv_disk}). The details of observations are given in Table \ref{uvit_obs}.  The image is flat-fielded, corrected for distortion and fixed pattern noise of the detector with the help of calibration files \citep{girish2017,postma2011}. The image has 4096$\times$4096 pixel dimension with 1 pixel corresponding to 0.4 $\arcsec$ , which is equivalent to 7.6 pc at the distance of IC 2574. All the calibration measurements of UVIT used in this study are adopted from the calibration paper of the instrument by \citet{tandon2017}. We also used the moment 0 H$~$I map, obtained from The H$~$I Nearby Galaxy Survey (THINGS) \citep{walter2008}, of the galaxy in our study.

\begin{table*}
\centering
\caption{Details of UVIT observations}
\label{uvit_obs}
\begin{tabular}{p{2cm}p{2cm}p{3cm}p{3cm}p{2cm}p{3cm}}
\hline
Filter & Bandpass & ZP magnitude & Unit conversion & $\triangle\lambda$ & Exposure time\\
 & ($\AA$) & (AB) & (erg/sec/cm$^2$/$\AA$) & ($\AA$) & (sec)\\\hline
F148W & 1250-1750 & 18.016 & 3.09$\times10^{-15}$ & 500 & 10375\\\hline
\end{tabular}
\end{table*}

\section{Extinction in UV}
\label{ext_s}
The nature of extinction in UV band shows characteristic variation for different external galaxies. \citet{gordon2003} studied the behaviour of extinction in the LMC, the SMC and the Milky way and found differences specifically in the UV regime. Since FUV radiation is sensitive to extinction, we considered both the Galactic extinction foreground to IC 2574 and interstellar extinction within the galaxy IC 2574 in our study. The E(B$-$V) values for Galactic foreground reddening and interstellar reddening of the galaxy IC 2574 are reported to be 0.036 and 0.013 mag respectively \citep{pasquali2008}. The nature of extinction law in IC 2574 was considered to be SMC type in the previous study by \citet{pasquali2008} and \citet{stewart2000}. In order to calculate extinction value in F148W filter, we considered Fitzpatrick law \citep{fitzpatrick1999} for foreground and SMC bar type extinction law \citep{gordon2003} for the interstellar reddening of the galaxy IC 2574. The extinction coefficients are calculated using \textit{extinction law calculator}\footnote{http://www.cadc-ccda.hia-iha.nrc-cnrc.gc.ca/community/YorkExtinctionSolver/coefficients.cgi} of \citet{mccall2004} available in NASA/IPAC Extragalactic Database (NED). We considered $R_{V}$ values as 3.07 and 2.75 respectively for Fitzpatrick and SMC bar type extinction law and estimated the average value of extinction coefficient ($R_{F148W}$) in F148W band (for the range 1250 - 1750 $\AA$). The estimated values of $R_{F148W}$ are found to be 8.57 and 13.06 respectively for Fitzpatrick and SMC bar type law. We calculated the value of extinction using equation 
\begin{equation}
\label{extinction}
A_{F148W} = R_{F148W}E(B-V),
\end{equation} 
where $R_{F148W}$ is the extinction coefficient, and found it as 0.31 mag and 1.70 mag respectively for foreground and interstellar extinction. The observed fluxes are corrected for both these extinctions in our analysis. It is to be noted that the value of interstellar extinction of the galaxy may have spatial variation and that can change the value of some of our estimated parameters.


\section{Analysis}
\label{analysis_s}

\begin{figure*}
\begin{center}
\includegraphics[width=6.5in]{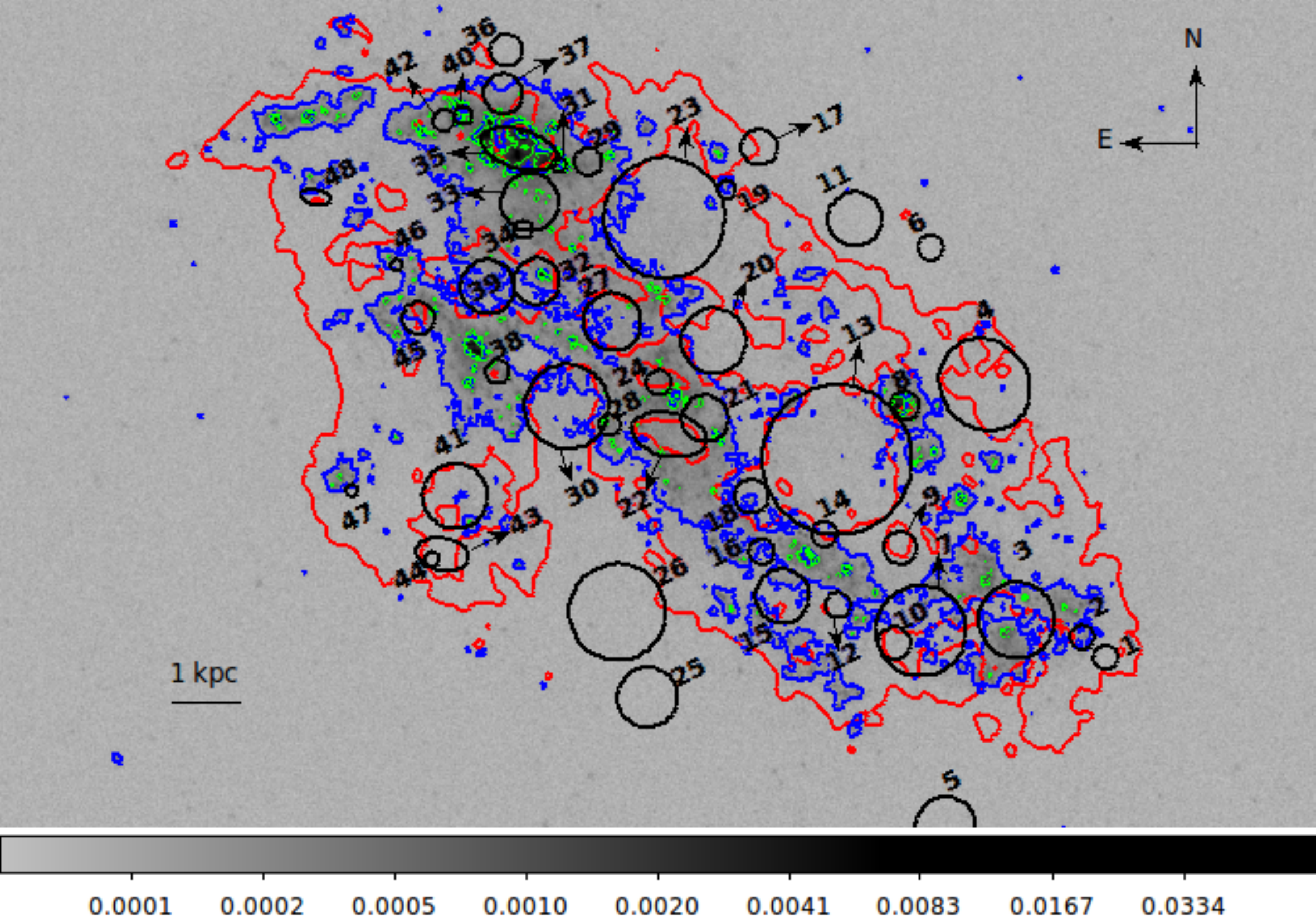} 
\caption{The background image is the UVIT FUV image of the galaxy IC 2574 observed in F148W filter. The H$~$I holes, as identified by \citet{walter1999}, are shown by black circles/ellipses. The number corresponding to each hole is same as assigned by these authors. The green and blue contours signify pixels with FUV flux more than $2.14\times10^{-18}$ and $3.21\times10^{-19}$ $erg/sec/cm^2/\AA$ respectively. The red contours denote regions with H$~$I column density greater than $1.0 \times 10^{21} cm^{-2}$.}
\label{shell}
 \end{center}
 \end{figure*}

\subsection{Distribution of FUV emission and H~I gas}
\label{fuv_distribution_s}
We considered the FUV image of the galaxy IC 2574 to locate young star forming regions. The distribution of FUV bright regions in the galaxy are mostly found to be clumpy in nature (Figure \ref{ic2574}). In order to understand the extent and characteristics of these regions more clearly, we generated contours for two different threshold fluxes and displayed them in Figure \ref{shell}. The blue and green contours shown in the figure respectively signify threshold value of $3.21\times10^{-19}$ and $2.14\times10^{-18}$ $erg/sec/cm^2/\AA$ , which are respectively equivalent to 3 and 20 times the average background flux. The reason behind choosing these two different values for generating the contours is to trace both the brighter and the fainter features together in the FUV emission profile of IC 2574. By choosing this lower limit, we expect to get rid of the background emission completely and to trace emission region only due to the galaxy. The green contours which trace relatively massive and more active star forming complexes, are found to be more compact in nature whereas we noticed the blue contours, which trace the overall FUV emission profile of IC 2574, to be more extended around the green contoured regions. Therefore, the star forming regions of the galaxy are found to have an overall patchy distribution with several compact clumps dispersed throughout the disk. \\\\
FUV emission, which mimics the current SFR in a galaxy, is found to correlate well with $H_{2}$ gas, whereas it does not follow a universal relationship with H$~$I gas throughout the galaxy \citep{bigiel2008}. In order to check the correlation of identified FUV bright regions with the column density of H~I gas, we plotted contours for $n_{HI} > 10^{21} cm^{-2}$ which is shown in red in Figure \ref{shell}. These density contours are generated from the VLA moment 0 H$~$I map of the galaxy available in \citet{walter2008}. The identified FUV bright star forming regions (both green and blue contours) of the galaxy are mostly found to be present within the extent of plotted H$~$I density contour. This simply conveys that star formation is mainly happening in regions with H~I column density greater than $10^{21} cm^{-2}$, which is known as the threshold value for star formation in IC 2574 \citep{walter1999}. There is no major star formation found outside the extent of the red contours.

\subsection{H$~$I shells}
The gaseous component of the galaxy IC 2574 has many hole like structures distributed all over its disk. Using VLA H$~$I observations, \citet{walter1999} identified 48 H$~$I shells and holes in the galaxy with radius in the range 100 - 1000 pc. They also found $H\alpha$ emission along the rims of larger holes which in other word signifies propagating star formation triggered due to the expansion of these holes. An expanding H$~$I hole can compress their surrounding ISM to trigger further star formation in a galaxy \citep{tagle2005}. In order to explore this, we have shown all these 48 holes as identified by \citet{walter1999}, overlaid on the FUV image of the galaxy in Figure \ref{shell}. A careful look reveals that many holes have bright FUV emissions (green contours) in their shells whereas some show signature of FUV emission inside the hole as well. We found 30 holes with FUV emission in their periphery while 15 holes show emission inside them. This signifies that star formation in certain parts of the galaxy could be triggered due to the expansion of H~I holes. It is also possible that some of these emissions are actually coming from other parts of the galaxy and they seem to be related to the holes due to projection in the sky plane.\\

As FUV radiation traces star formation up to a few hundred Myr and the H~I holes of IC~2574 are of $\le $ 50 Myr \citep{walter1999}, a detection of FUV bright region is not sufficient to confirm the existence of hole driven star formation. In order to verify whether the FUV regions are actually young (age $<$10 Myr), it is important to check whether they also show H$\alpha$ emission. We used the catalog of H$\alpha$ emitting regions of \citet{miller1994} for cross-identification. About 54\% of the FUV bright regions, detected in the shells, are also found to have H$\alpha$ emission. These are potential locations where star formation could be triggered due to the expanding holes. The emission identified inside the holes (specifically with H$\alpha$ counter part) can actually be coming from the shell, present either in front or back along the line of sight. We discuss these in detail in the next section. It is further noticed that some of the green contours in different places across the galaxy are present in between multiple holes. This gives an appearance of star formation happening in regions which are surrounded by holes. The feedback from the expanding holes may have compressed the ISM in between to make those sites more favourable for star formation. Both the above scenarios signify the impact of hole expansion in triggering star formation inside the galaxy. \\\\
Again, the fact that star forming regions are mostly identified in regions with H$~$I density more than $10^{21} cm^{-2}$ signifies that the mechanism which has shaped the H$~$I distribution inside the galaxy has also triggered secondary star formation in some of those sites. Therefore, in majority of the cases, the FUV emission, H$~$I density and the shell structure are likely to be co-spatial.

\subsection{FUV bright star forming regions}
\label{fuv_clump_s}

\begin{figure*}
\centering
\includegraphics[width=7.0in]{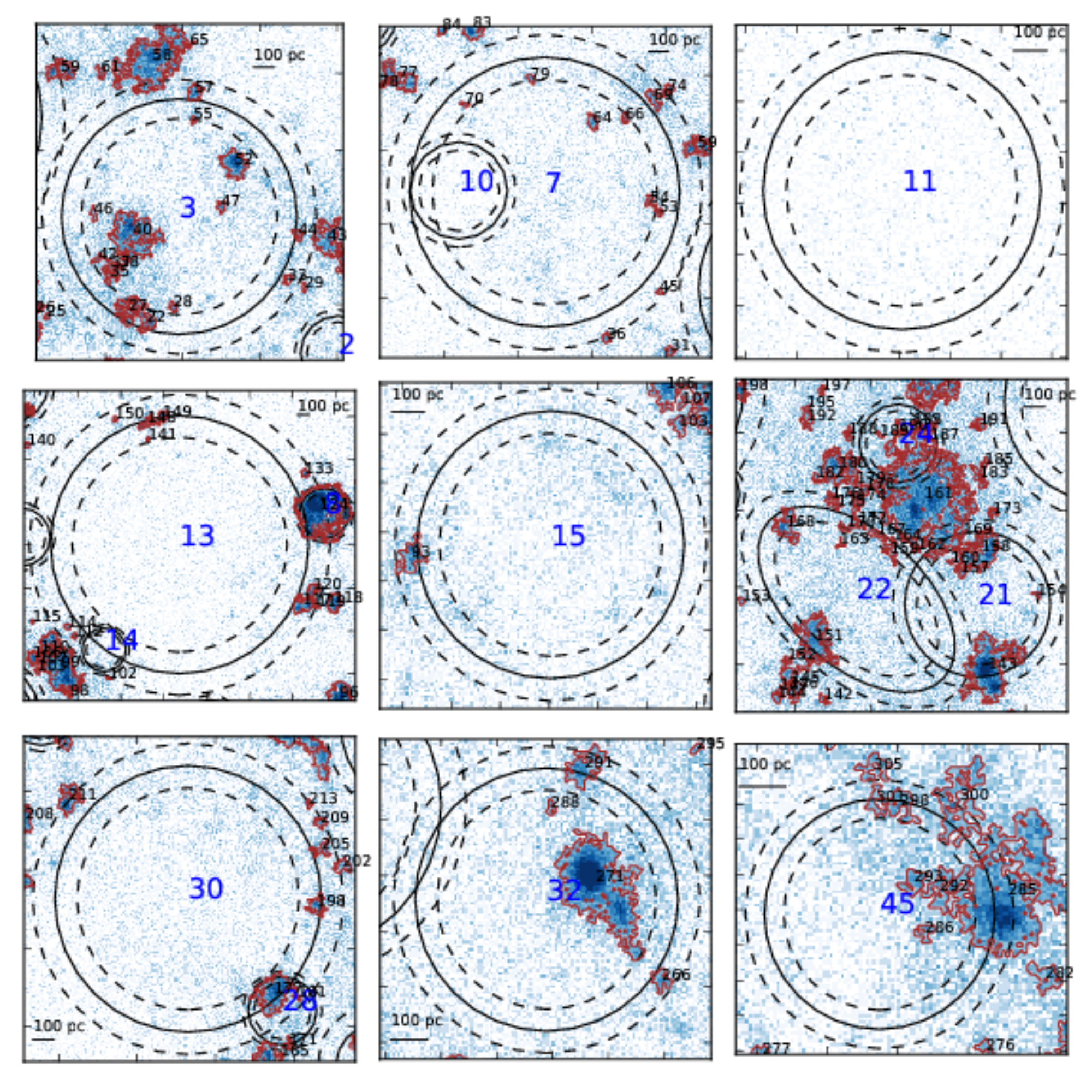} 
\caption{The figure shows some selected holes along with the identified FUV bright regions. The background image is UVIT FUV image of IC 2574. The black solid circles/ellipses show the holes of radius R, where the shells are defined by black dashed lines of width R/3 between radii (R$-$R/6) to (R$+$R/6). A length scale of 100 pc is shown in each image. The brown contours denote the FUV bright regions identified for a threshold flux of $1.07\times10^{-18}$ $erg/sec/cm^2/\AA$. The numbers shown in blue and black respectively signify the ID for holes and identified regions as given in Table \ref{clumps_remark}.}
\label{uv_clumps}

\end{figure*}
 
In order to identify FUV bright star forming regions across the galaxy, we used the Python package \textit{astrodendro}\footnote{http://www.dendrograms.org/}. This package helps to identify structures formed by a minimum number of pixels each having a value more than a defined threshold flux. It identifies both parent (larger structure) and child structures (smaller sub-structures within the larger one) on the basis of selected parameter value. We considered the threshold flux as $1.07\times10^{-18} erg/sec/cm^2/\AA$, which is 10 times the average FUV background flux. This threshold corresponds to the flux of a B5 spectral type star at the distance of IC 2574. The minimum number of pixels for the identification of regions is considered as 10, which corresponds to a radius of $\sim$ 15 pc for a circular area. The value of this lower limit is fixed by balancing two facts. One is to resolve the smaller clumps well and other is to avoid identifying a lot of them. With these parameters, we identified a total of 403 parent structures in the galaxy within a galactocentric distance of 10 kpc. These FUV bright regions are likely to be the active star forming regions of the galaxy. Out of the 403 parent clumps, 96 (23.8\%) are found to have multiple smaller sub-structures within them whereas the rest 307 (76.2\%) clumps show no sub-structures within the detection threshold. Among all the identified parent structures, we found one, located in the north-eastern part of the galaxy, to be exceptionally larger than the rest. This parent structure contains many smaller sub-structures inside it. We selected 17 bright sub-structures instead of this single large structure for our study. Hence we have a total of 419 FUV bright regions, which we considered for the rest of our analysis.\\

In order to identify an FUV bright region to be formed due to an expanding H~I shell, we need to define a thickness to the shell and check for FUV bright regions located within this shell. Here, we considered the width of the shell as R/3 (where R is the radius of the hole), from radii (R$-$R/6) to (R$+$R/6) for each hole (Figure \ref{uv_clumps}). Implication of this assumption is discussed in section \ref{discussion_s}. If an FUV bright region (or some part of it) is present within the radii (R$-$R/6) to (R+R/6) of any hole, then it is considered as part of the shell of that hole. If it is present within radius (R$-$R/6), then we assume it to be inside the hole. In case the region is located at a distance more than (R$+$R/6) from the hole centre, we considered it to be not related to that hole. Following this methodology, we identified regions which are connected with holes (either in the shell or inside) and listed them in Table \ref{clumps_remark}. \citet{walter1999} estimated the ages of these 48 holes and also classified them in three different types (Type 1, 2, 3) on the basis of their appearance in the p-V diagram. A hole is defined as Type 1, when neither of the receding and approaching sides of the hole can be observed in p-V diagram. In the case, when p-V diagram of a hole shows deformation indicating that it is offset with respect to galaxy plane is defined as Type 2 hole. A hole is called Type 3 when the velocity of both the receding and approaching sides can be measured. We noted the information of hole types in Table \ref{clumps_remark} from \citet{walter1999}.\\

Out of 419 identified regions, we found 120 (28.6\%) to be present in shells and 53 (12.6\%) to be present inside holes. It is to be noted that the list is made as per their locations in the projected sky plane and hence there can be a possibility to have a slightly different scenario in the galaxy plane. We also found 252 (60.1\%) regions as not related to any of the holes. Therefore, 60.1\% of the star formation happening in the galaxy has no connection with the holes. In other words, 30 (62.5\%) out of 48 holes are found to have FUV emission in their shell whereas 15 (31.2\%) holes show emission inside. We noticed 16 (33.3\%) holes to have no related FUV emission. There are 13 holes which show FUV emission both in shell and inside them.\\

In Figure \ref{uv_clumps}, we have shown some selected holes of various types along with the identified FUV bright regions. In the case of Hole 3 (top-left of Figure \ref{uv_clumps}), we noticed 8 regions (Regions 22, 27, 33, 43, 44, 55, 57, 58) to be present in the shell and 9 regions (Regions 28, 35, 37, 38, 40, 42, 46, 47, 52) are identified inside it (Table \ref{clumps_remark}). The regions found inside can actually be present in the shell but it appears to be inside the hole only in the projected plane. The other holes (except Hole 11) are also found to have emissions in their shell. Hole 11 is an example of holes which do not have any related FUV emission. Hole 8 and 32 show strong FUV emission inside them. We also noticed FUV emission in region between Hole 21, 22 and 24. This is an example where star formation is possibly triggered due to expansion/collision of multiple holes.\\

We further checked the H$\alpha$ counter part for each FUV region from the catalog of \citet{miller1994}. In case there is a cross-match, we mentioned the ID number of the H$\alpha$ emitting region (from their catalog) for the corresponding FUV region in parenthesis in Table \ref{clumps_remark}. We noticed 65 (15.5\% of total) out of 120 regions identified in the shell to have H$\alpha$ emission also. The detection of H$\alpha$ emission signifies a recent star forming activity in these regions. Therefore, these are the most possible sites where star formation has been triggered recently due to expanding H~I holes. We found 16 out of 53 regions present inside the holes to have H$\alpha$ detection. These 16 regions therefore have recent star formation and hence it is more possible that they are actually present in shells, but appears to be located inside holes. The same exercise is also performed for the 252 non-related regions and we noticed 106 of them to have H$\alpha$ counter part. In Table \ref{table3_sum}, we have extracted the summary of Table \ref{clumps_remark} for different types of hole and the identified FUV regions. We found that the number of holes with emission identified in their shells are 12, 5 and 13 respectively for Type 1, Type 2 and Type 3 holes respectively. Among the 15 holes with emission inside, we found 9 of Type 3, 4 of Type 1 and 2 of Type 2. It is noticed that 23 out of 30 holes, which show FUV emission in their shell, also have $H\alpha$ emission. Therefore, 47.9\% of the holes show positive signature of triggered star formation in their shell. We did not find any significant correlation between holes with/without triggered star formation and their ages.\\

To understand the physical properties of these 419 identified regions, we estimated several parameters and listed in Table \ref{uvit_clumps_table}. 
We estimated the radius (R) in pc and the galactocentric distance ($R_{gc}$) in kpc for each region. The radius, which is estimated by equating the area of the irregular shaped regions to that of a circle, is found to have a range between $\sim$ 15 - 285 pc. 95\% of our identified regions have radii smaller than 100 pc. The galactocentric distances are estimated by assuming the galaxy centre, inclination and position angle of IC 2574 from Table \ref{ic2574} and using the relation given in section 2 of \citet{marel2001}.\\\\ 

\subsubsection{Estimation of FUV flux, Luminosity and SFR}
The \textit{astrodendro} package, discussed in section \ref{fuv_clump_s}, also provides area and flux for the identified regions. Considering these two, we measured the background and extinction corrected FUV flux ($erg/sec/cm^2/\AA$) for the regions.
The background flux is estimated from the average flux measured in four different circular regions of radius 1 arcmin present in the galaxy field. We choose the radius to be 1 arcmin as it is around 145 pixels in the image and thus covers a good area for estimating the average background. The circle of 1 arcmin radius also fits well in the space between the edge of the detector and the extent of the galaxy. The corrected fluxes are used to estimate luminosity density ($erg/sec/pc^2$) and SFR density ($M_{\odot}/yr/kpc^2$) corresponding to each region. We used the relation given by \citet{kara2013} for estimating the SFR from the measured FUV magnitude. The relation is given in 
\begin{equation}
log(SFR_{FUV} (M_{\odot}/yr)) = 2.78 - 0.4*mag_{FUV} + 2log(D), 
\label{sfr_eq}
\end{equation}
where $mag_{FUV}$ is the apparent FUV magnitude (AB system) and D is the distance to the galaxy in Mpc. The measured value of SFR density shows a range between 0.0238 - 0.5409 $M_{\odot}/yr/kpc^2$ which highlights the diversity of star forming regions in the galaxy. \\

\subsubsection{Estimation of H~I density}
\begin{figure}
\begin{center}
\includegraphics[width=3.3in]{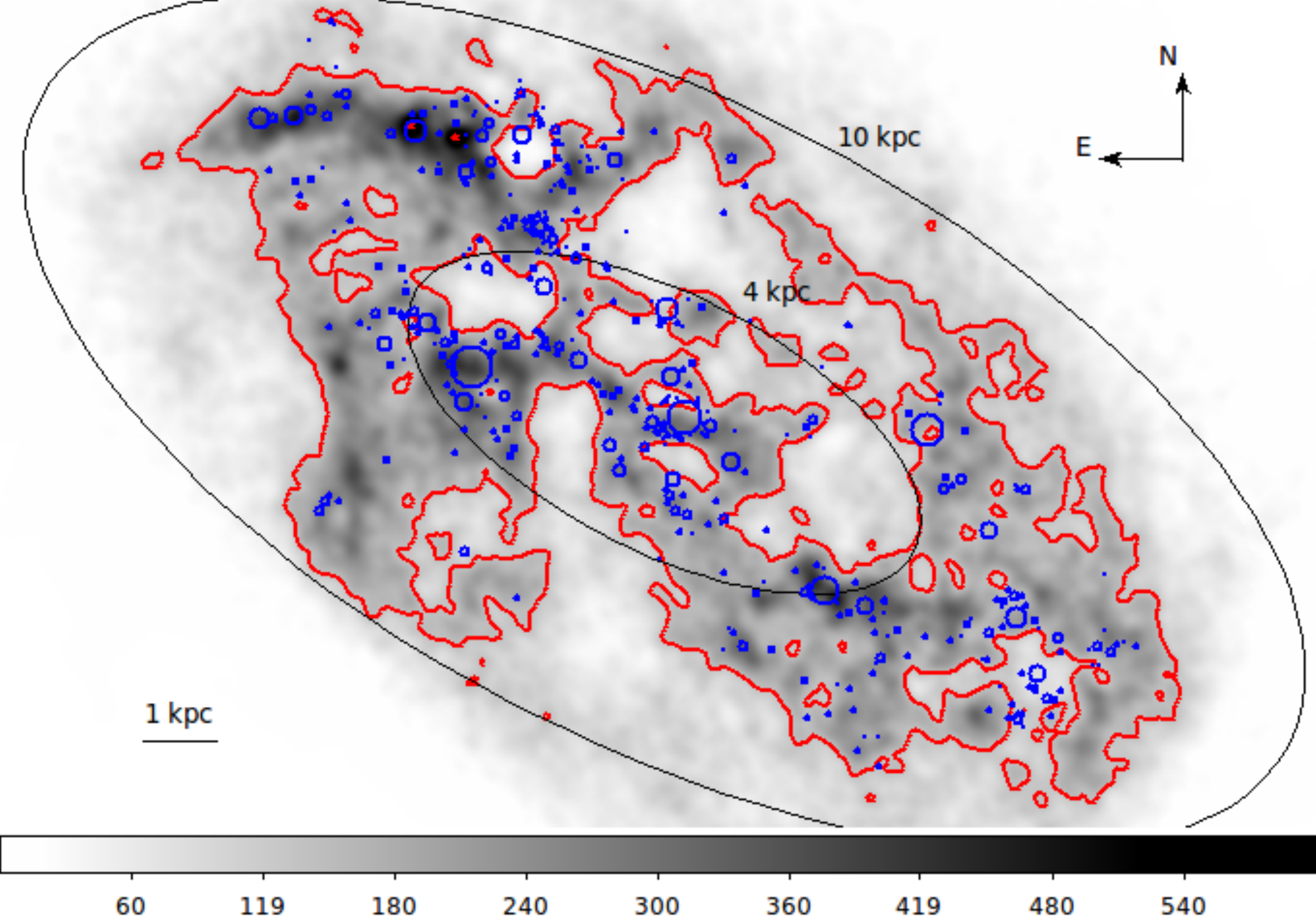} 
\caption{The background image is H$~$I moment 0 map of IC 2574. The gray scale signifies flux in JY/B*M/S. The red contours signify regions having H$~$I column density more than $10^{21} cm^{-2}$. The identified FUV bright star forming regions are shown as blue circles.}
\label{uv_clumps_hi}
\end{center}
\end{figure}

\begin{figure}
\begin{center}
\includegraphics[width=3.7in]{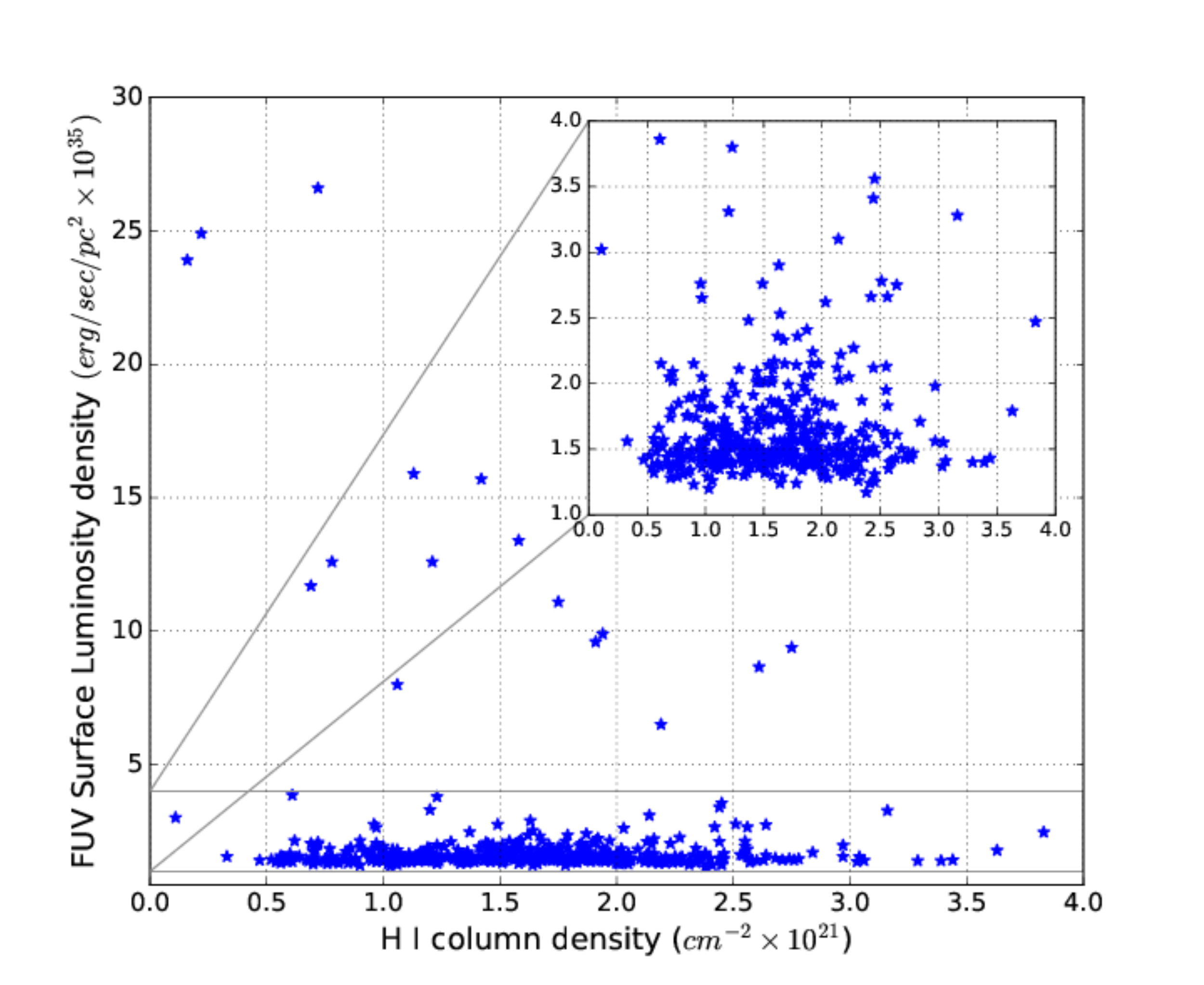} 
\caption{The measured value of FUV surface luminosity density and H$~$I column density of 419 identified FUV bright regions are shown. The inset shows the zoomed in part for FUV luminosity density between $1 - 5\times10^{35}$ $erg/sec/pc^2$.}
\label{cps_uv_hi}
\end{center}
\end{figure}

As the estimated values of SFR density show a wide range, it will be interesting to correlate it with the average H~I column density of these regions. In Figure \ref{uv_clumps_hi}, we have shown the H$~$I column density map of the galaxy from \citet{walter2008} along with 419 FUV bright regions (blue circles). These circles are plotted for the equivalent area of each identified region. The red contours signify regions with H$~$I column density greater than $10^{21}  cm^{-2}$. The identified star forming regions are mostly found to be present in locations with dense H$~$I gas. We measured the average column density of neutral hydrogen for the FUV bright regions from the moment 0 H$~$I map. The estimated values show a range between 0.11 - 3.83 $\times10^{21}  cm^{-2}$. In Figure \ref{cps_uv_hi}, we have shown the FUV luminosity density and H$~$I column density for all these 419 regions. The average H~I column density is found to be more than $10^{21} cm^{-2}$ for 345 ($\sim82.3\%$) regions. Out of these 345 regions, 241 have values between $10^{21} cm^{-2}$ and $2\times 10^{21} cm^{-2}$, 95 between $2\times10^{21} cm^{-2}$ and $3 \times 10^{21} cm^{-2}$ and 9 with value greater than $3\times 10^{21} cm^{-2}$. The rest 74 regions ($\sim17.7\%$) show column density less than $10^{21} cm^{-2}$. The estimated value of the H~I column density will differ slightly from the actual value due to the circular approximation of the regions. Similarly for FUV surface luminosity density, we found 347 regions ($\sim82.8\%$) to have value between $10^{35}$ and $2\times10^{35}$ $erg/sec/pc^2$ and 72 regions ($\sim17.2\%$) have value more than $2\times10^{35}$ $erg/sec/pc^2$ with 11 of them brighter than $10\times10^{35}$ $erg/sec/pc^2$.

\begin{deluxetable*}{ccccc}
\centering
\tablewidth{17.0cm}

\caption{Connection between FUV bright star forming regions and the H$~$I holes. This table helps to capture the location of the FUV bright regions with respect to the H~I holes. In the second column we listed the regions present in the shell. The regions present inside each hole are shown in column 3. Each region number signifies its ID, listed in Table \ref{uvit_clumps_table}. The numbers given within parentheses are the ID of H$\alpha$ cross-identified region from the catalog of \citet{miller1994}. The age of each hole and their types are given in columns 4 and 5 respectively from \citet{walter1999}.}

\label{clumps_remark}
\startdata \\
\hline
Hole & Region in shell & Region inside hole & Age of hole (Myr)\tablenotemark{a} & Hole type\tablenotemark{a}\\\hline
1 & -- & -- & 19.6 & 3\\
2 & -- & -- & 31.3 & 3\\
3 & 22,27,33,43(3),44,55,57(5),58(9) & 28,35,37,38,40(6),42,46,47,52(4) & 29 & 3\\
4 & 123(24) & -- & 42.5 & 2\\
5 & -- & -- & 43.5 & 3\\
6 & -- & -- & 20.6 & 3\\
7 & 36(18),45,53,59(16),69,70,74(22),79(42) & 54,64,66 & -- & 1\\
8 & -- & 124(32,34,36,38,39,40,41,44) & 12.2 & 3\\
9 & -- & -- & 30.2 & 3\\
10 & -- & -- & -- & 1\\
11 & -- & -- & 44.1 & 3\\
12 & -- & -- & -- & 1\\
13 & 99(57,58,60,63),102,112(61),114,117(30,33), & -- & -- & 1\\
 & 119,120,124(32,34,36,38,39,40,41,44), & & & \\
 & 133(45),141,148(66),149(64),150 & & & \\
14 & -- & -- & -- & 1\\
15 & 93 & -- & 58.7 & 3\\
16 & 116 & -- & 20.6 & 3\\
17 & -- & -- & 18.1 & 3\\
18 & -- & 122 & 19.9 & 3\\
19 & 281 & -- & 14.7 & 3\\
20 & 199(92) & -- & -- & 1\\
21 & 143(80,83,84),154(78),161(94,96,102,104),162,169 & 157(89),158(87,90),160(89) & 17 & 3\\
22 & 151(103),159,162,164,167,168,170,172, & 163 & 28.9 & 3\\
 & 174(107),175,176 & & & \\
23 & 232,235(109),241(109),330 & 269,287 & -- & 1\\
24 & 161(94,96,102,104) & 187,189,190,193 & -- & 1\\
25 & -- & -- & 30.1 & 3\\
26 & -- & -- & -- & 1\\
27 & 231,249 & -- & -- & 1\\
28 & 171(115) & 177(121),181(121) & 23.5 & 3\\
29 & 343(118),356(128),358(128) & 346,349,354 & 18 & 3\\
30 & 177(121),181(121),198(120),202(119),205, & -- & -- & 1\\
 & 209(124),211(173),213 & & & \\
31 & 1(133),2(134) & -- & -- & 1\\
32 & 266(136),291(165) & 271(143,155),288(155) & 12.7 & 3\\
33 & 303(144),308,310(151),311,313(144),314, & 318(150),319,320(150),322,327,328,329, & 16.3 & 2\\
 & 323,335,340(186),344,348,350,360 & 333,334,337,338,341,355(178) & & \\
34 & 319,327,328 & -- & 13.1 & 3\\
35 & 7(137),8(137),9(138),10(138),14(192),15(194), & 5(148),6(148),12(172),13 & 14.3 & 2\\
 & 16(198,202),17(167),364,365 & & & \\
36 & -- & -- & 21.1 & 3\\
37 & 386(170),393,396 & 387,389 & -- & 1\\
\hline
\enddata 
\end{deluxetable*} 

\setcounter{table}{2}
\begin{deluxetable*}{ccccc}
\centering
\tablecaption{continued}
\startdata \\
Hole & Region in shell & Region inside hole & Age of hole (Myr) & Hole type\\\hline
38 & 226(181),233(193,197,199,203,206,208,218), & -- & -- & 1\\
 & 234,238 & & & \\
39 & 263(184),304,306,321 & -- & -- & 1\\
40 & 3(224) & -- & 11.7 & 3\\
41 & 156(212) & -- & -- & 1\\
42 & 385(235),391(235),403 & -- & 10.8 & 2\\
43 & -- & -- & -- & 2\\
44 & -- & -- & 11.7 & 3\\
45 & 285(236),298,300(241),301,305 & 286,292,293 & 30.2 & 3\\
46 & -- & -- & 19.6 & 3\\
47 & 220 & -- & 9.8 & 3\\
48 & 394 & -- & 17.6 & 2\\
\hline
\enddata
\tablenotetext{a}{From the study of \citet{walter1999}}

\end{deluxetable*}

\begin{table*}
\centering
\caption{Summary of Table \ref{clumps_remark}. Column 2 shows the total number for holes (along with three different types) and identified regions. Columns 3 and 4 respectively show the number of holes with FUV emission in shell and inside it. Column 5 lists the number of holes with no related FUV emission. The bottom row of the table denotes the similar statistics with respect to the identified regions. The numbers shown in parenthesis denote the numbers after cross-match with H$\alpha$ emission.}
\label{table3_sum}
\begin{tabular}{p{3cm}p{3cm}p{4cm}p{4cm}p{3cm}}
\hline
Hole & Total number & FUV emission in shell & FUV emission inside hole & With no related FUV emission\\\hline
Total hole & 48 & 30(23) & 15(7) & 16\\
Type 1 hole & 16 & 12(11) & 4(0) & 4\\
Type 2 hole & 6 & 5(4) & 2(2) & 1\\
Type 3 hole & 26 & 13(8) & 9(5) & 11\\
\hline
 & Total number & Present in shell & Present inside hole & Not related with hole\\\hline 
FUV Region & 419 & 120(65) & 53(15) & 252(106)\\\hline

\end{tabular}
\end{table*}

\begin{table*}

\centering
\caption{Properties of FUV bright star forming regions as defined in Figure \ref{uv_clumps}. The full table containing all 419 regions is available in electronic format.}
\label{uvit_clumps_table}
\begin{tabular}{p{1.0cm}p{1.7cm}p{2.0cm}p{1.2cm}p{1cm}p{0.7cm}p{1.4cm}p{1.2cm}p{1.7cm}}
\hline
  Region & RA (J2000) (hh:mm:ss.s) & DEC (J2000) (dd:mm:ss.s) & Radius (pc) & $R_{gc}$\tablenotemark{a} (kpc) & Flux\tablenotemark{b} & Luminosity density\tablenotemark{c} & SFR density\tablenotemark{d} & Average HI column density\tablenotemark{e}\\\hline
1 & 10:28:37.1 & +68:27:57.1 & 22.2 & 5.64 & 1.70 & 9.39 & 0.1910 & 2.75\\
2 & 10:28:37.2 & +68:28:1.9 & 27.4 & 5.79 & 2.38 & 8.66 & 0.1762 & 2.61\\
3 & 10:28:53.0 & +68:28:35.1 & 29.6 & 6.25 & 2.10 & 6.51 & 0.1324 & 2.19\\
4 & 10:28:53.4 & +68:28:49.8 & 33.9 & 6.67 & 4.70 & 11.13 & 0.2263 & 1.75\\
5 & 10:28:40.5 & +68:28:1.0 & 22.6 & 5.58 & 2.37 & 12.61 & 0.2564 & 0.78\\
6 & 10:28:40.1 & +68:28:3.6 & 25.3 & 5.69 & 2.76 & 11.73 & 0.2386 & 0.69\\
7 & 10:28:38.8 & +68:28:6.7 & 17.1 & 5.86 & 1.35 & 12.56 & 0.2554 & 1.21\\
8 & 10:28:38.3 & +68:28:6.7 & 19.1 & 5.89 & 1.80 & 13.41 & 0.2727 & 1.58\\
9 & 10:28:38.5 & +68:28:8.9 & 14.8 & 5.95 & 1.27 & 15.73 & 0.3199 & 1.42\\
10 & 10:28:38.9 & +68:28:9.4 & 16.6 & 5.94 & 1.60 & 15.92 & 0.3238 & 1.13\\
11 & 10:28:39.0 & +68:28:30.3 & 44.9 & 6.63 & 5.91 & 8.00 & 0.1627 & 1.06\\
12 & 10:28:44.5 & +68:28:10.4 & 24.9 & 5.72 & 5.46 & 23.93 & 0.4868 & 0.16\\
13 & 10:28:44.4 & +68:28:7.0 & 30.5 & 5.62 & 8.52 & 24.88 & 0.5061 & 0.22\\
14 & 10:28:48.1 & +68:28:4.2 & 66.9 & 5.43 & 15.79 & 9.60 & 0.1953 & 1.91\\
15 & 10:28:48.6 & +68:28:35.0 & 66.0 & 6.35 & 15.82 & 9.90 & 0.2014 & 1.94\\
16 & 10:28:49.2 & +68:28:25.0 & 87.6 & 6.03 & 46.57 & 16.52 & 0.3361 & 2.19\\
17 & 10:28:43.8 & +68:28:26.3 & 121.2 & 6.25 & 143.31 & 26.59 & 0.5409 & 0.72\\
18 & 10:27:21.8 & +68:18:43.5 & 32.0 & 9.79 & 1.14 & 3.02 & 0.0615 & 0.11\\
19 & 10:27:28.7 & +68:20:56.3 & 29.6 & 6.97 & 0.52 & 1.63 & 0.0331 & 1.15\\
20 & 10:27:32.8 & +68:20:48.6 & 18.6 & 6.81 & 0.22 & 1.76 & 0.0359 & 0.85\\
21 & 10:27:33.3 & +68:20:54.1 & 53.2 & 6.70 & 1.98 & 1.90 & 0.0387 & 0.90\\
22 & 10:27:28.3 & +68:21:10.1 & 41.9 & 6.84 & 1.16 & 1.80 & 0.0366 & 0.71\\
23 & 10:27:34.8 & +68:20:54.5 & 32.9 & 6.60 & 0.62 & 1.57 & 0.0320 & 1.01\\
24 & 10:27:34.1 & +68:20:55.3 & 21.0 & 6.63 & 0.24 & 1.49 & 0.0302 & 0.88\\
25 & 10:27:32.8 & +68:20:59.5 & 13.5 & 6.64 & 0.10 & 1.56 & 0.0318 & 1.00\\
26 & 10:27:33.4 & +68:20:59.5 & 26.0 & 6.60 & 0.35 & 1.41 & 0.0287 & 0.95\\
27 & 10:27:29.4 & +68:21:10.7 & 53.6 & 6.75 & 1.66 & 1.57 & 0.0320 & 0.64\\
28 & 10:27:27.5 & +68:21:16.6 & 21.0 & 6.85 & 0.27 & 1.66 & 0.0338 & 0.60\\
29 & 10:27:22.1 & +68:21:36.9 & 15.4 & 7.27 & 0.13 & 1.47 & 0.0298 & 1.08\\
30 & 10:27:17.0 & +68:21:51.4 & 21.0 & 7.87 & 0.34 & 2.14 & 0.0435 & 1.78\\
31 & 10:27:34.2 & +68:21:6.2 & 21.8 & 6.45 & 0.32 & 1.82 & 0.0370 & 0.95\\
32 & 10:27:20.4 & +68:21:46.7 & 54.4 & 7.45 & 1.97 & 1.82 & 0.0369 & 1.66\\
33 & 10:27:23.0 & +68:21:36.8 & 21.0 & 7.17 & 0.25 & 1.53 & 0.0311 & 0.98\\
34 & 10:27:53.1 & +68:20:18.4 & 22.6 & 7.24 & 0.36 & 1.89 & 0.0385 & 0.97\\
35 & 10:27:30.9 & +68:21:16.8 & 29.9 & 6.55 & 0.49 & 1.48 & 0.0301 & 0.58\\
36 & 10:27:37.5 & +68:21:1.6 & 22.2 & 6.34 & 0.28 & 1.56 & 0.0317 & 1.64\\
37 & 10:27:31.1 & +68:21:19.4 & 22.2 & 6.51 & 0.29 & 1.58 & 0.0320 & 0.56\\
38 & 10:27:30.7 & +68:21:20.3 & 14.8 & 6.53 & 0.11 & 1.39 & 0.0283 & 0.59\\
39 & 10:27:19.5 & +68:21:52.9 & 31.4 & 7.56 & 0.57 & 1.57 & 0.0319 & 1.76\\
40 & 10:27:30.8 & +68:21:29.6 & 104.5 & 6.43 & 8.63 & 2.15 & 0.0438 & 0.62\\\hline

\end{tabular}
\tablenotetext{a}{Galactocentric distance}
\tablenotetext{b}{Total FUV flux in $erg/sec/cm^2/\AA$ $\times 10^{-15}$}
\tablenotetext{c}{FUV Luminosity in $erg/sec/pc^2$ $\times 10^{35}$.}
\tablenotetext{d}{SFR is in $M_{\odot}/yr/kpc^2$.}
\tablenotetext{e}{Density is in $cm^{-2} \times 10^{21}$}

\end{table*}

\begin{figure*}
\begin{center}
\includegraphics[width=6.7in]{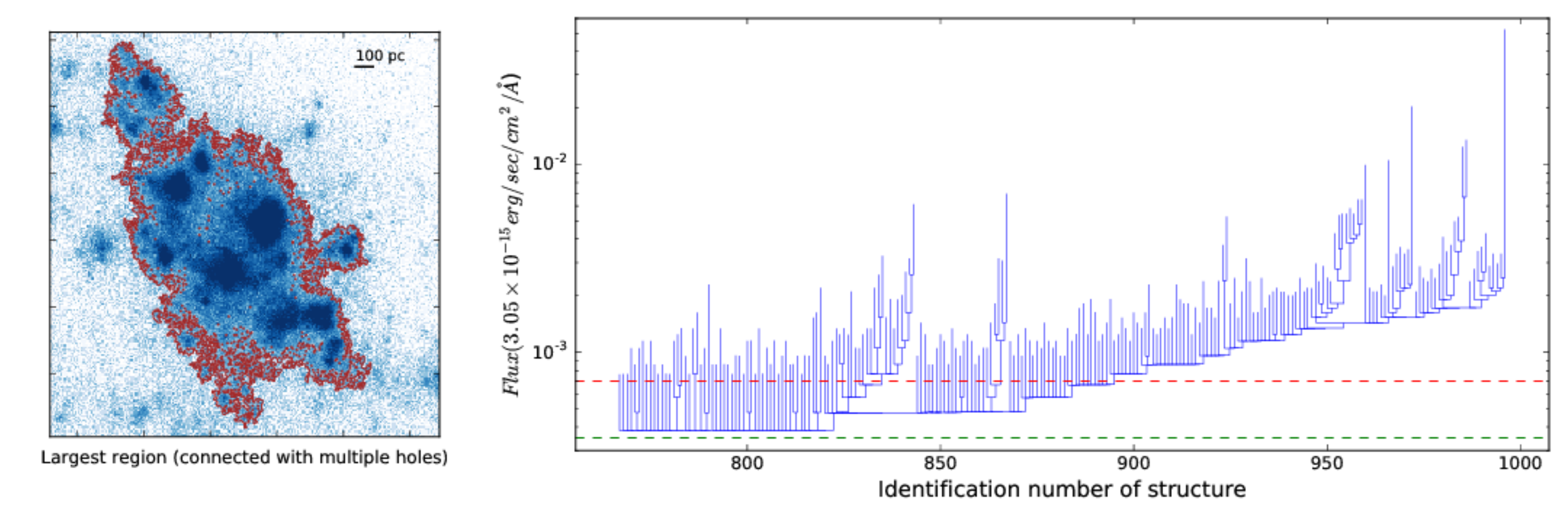}\\
\includegraphics[width=6.7in]{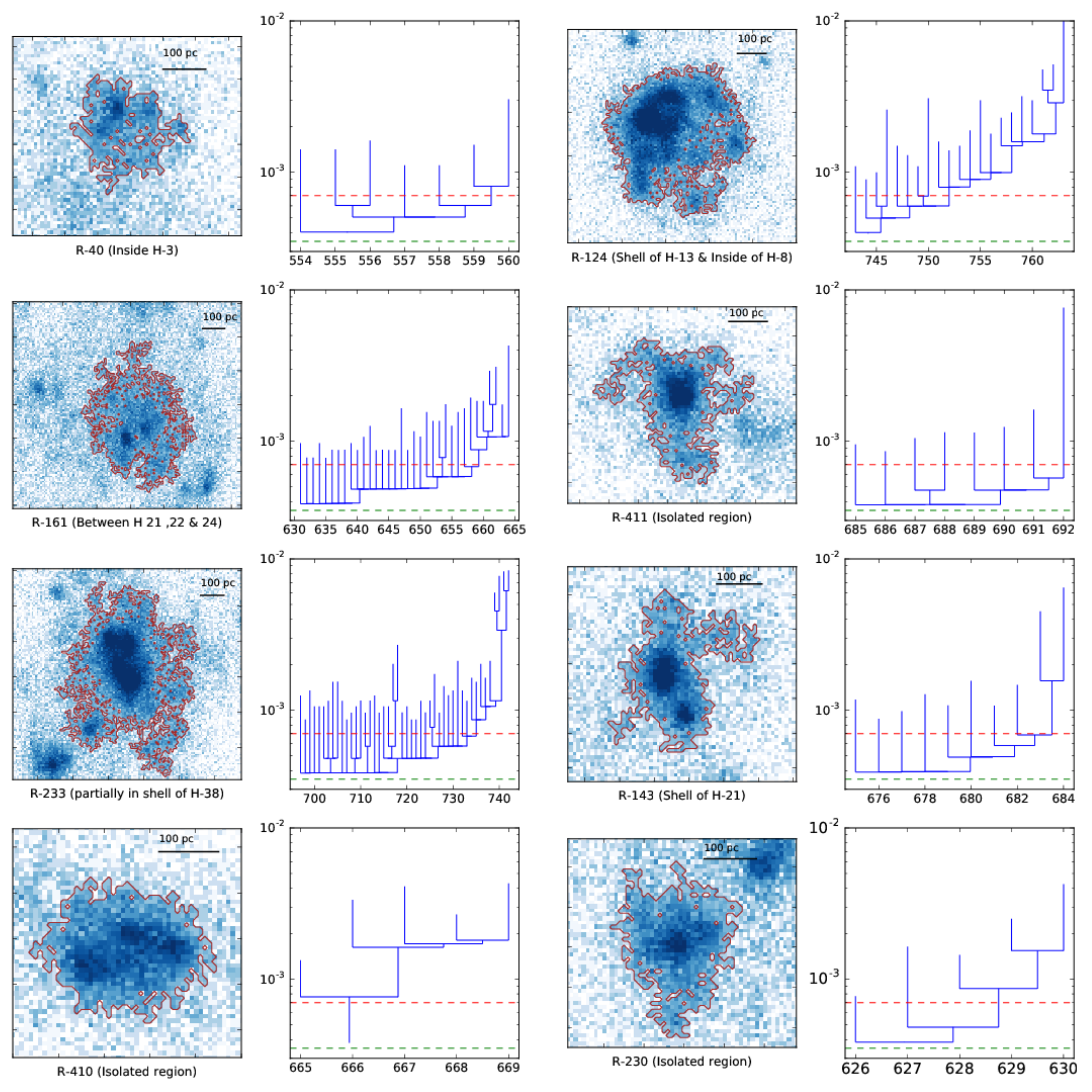}
\caption{The figure shows 9 selected parent structures, along with their dendrograms, identified for a threshold flux $1.07\times10^{-18} erg/sec/cm^2/\AA$ (green dashed line). The red dashed lines represent flux value of $2.14\times10^{-18} erg/sec/cm^2/\AA$. The largest parent structure and its dendrogram are shown in the upper panel of the figure. A length scale of 100 pc is shown for each region in solid black line. The Y axis of dendrogram shows the flux in terms of counts per second while the X axis denotes unique identification number of each structure which is not related to their ID presented in the paper.}
\label{dendro_all}
\end{center}
\end{figure*}

\subsection{Structure of star forming regions}
Since the structure of star forming regions in galaxies is known to be hierarchical in nature \citep{elmegreen2000}, we further explored the structural characteristics of identified star forming regions as a function of varying flux level. We selected 9 relatively larger parent structures from the list of our identified regions (Section \ref{fuv_clump_s}) and shown them in Figure \ref{dendro_all}. Among these, we have also included the largest parent structure (top panel of the figure) identified in the north-eastern part of the galaxy. Each of these parent structure contains multiple sub-structures of different size and flux level inside them. In order to explore the characteristics of sub-structures we have shown dendrogram for each region in the same figure. Dendrograms are structure trees used to highlight the parent-child connection between identified structures of different flux levels. Each tree present in Figure \ref{dendro_all} signifies a parent structure whereas the leaves connected to that tree represent child structures present within the parent structure. The selected regions have different significance as per their location in the galaxy. They are located either in shell, inside hole, in between multiple holes or away from any hole. The idea is to check the structural nature of star forming regions located in different environments of the galaxy. In the dendrograms, the y axis denotes the FUV flux level of each sub-structure. The dendrogram of the largest parent structure (top panel of the Figure \ref{dendro_all}), which is distinctly different from the rest, shows many leaves with different flux levels. The brightest star forming knots present in the galaxy are actually part of this large region. In case of other parent structures which are much smaller than this large region, we noticed similar nature of dendrogram with much reduced number of leaves, but with different flux levels. The brighter star forming clumps (sub-structure) identified throughout the galaxy are found to be present inside such large complexes (parent structure). Therefore, irrespective of location, the larger star forming complexes have multiple sub-structures inside them.

\begin{figure}
\begin{center}
\includegraphics[width=3.7in]{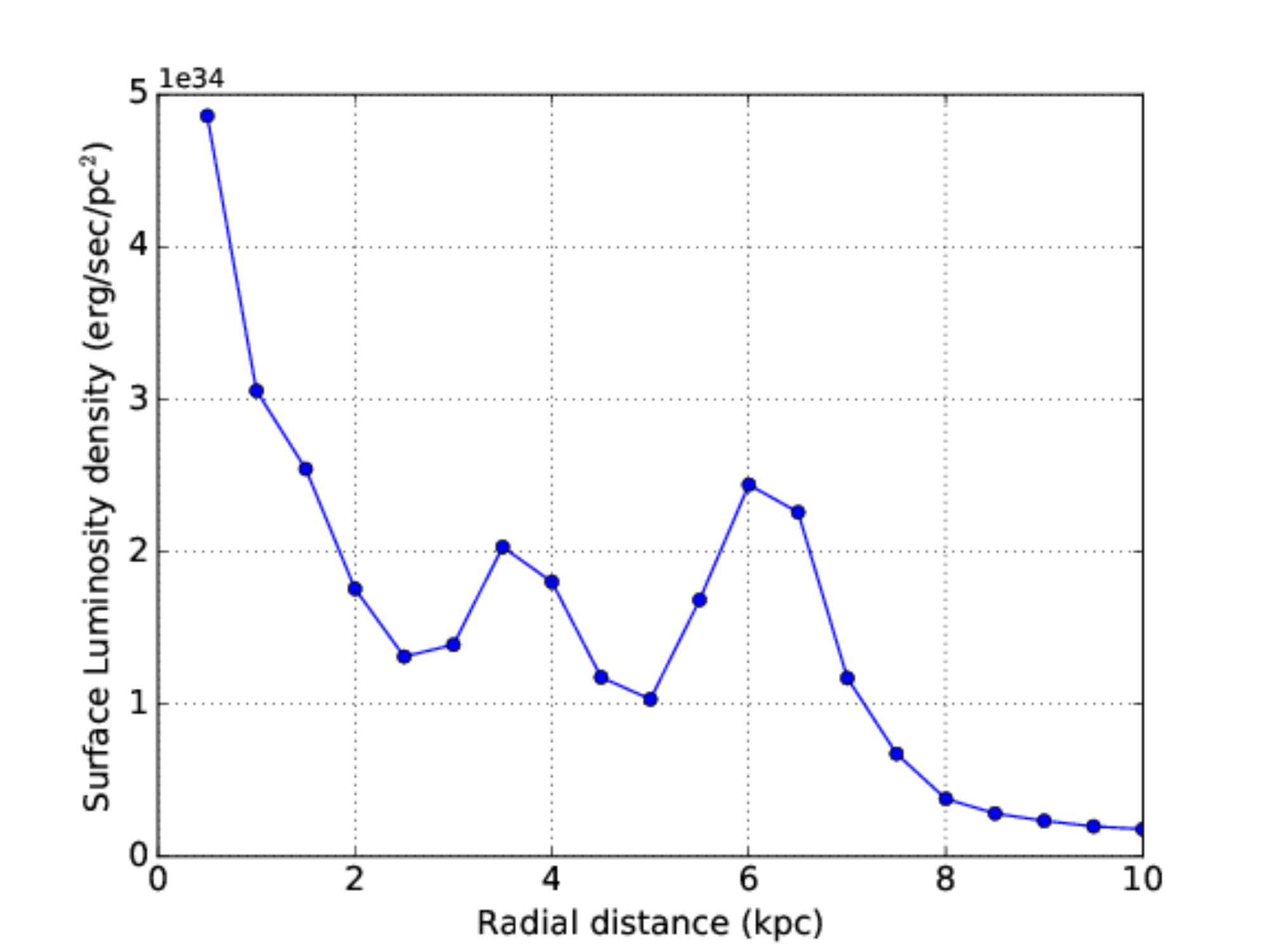} 
\caption{The radial surface luminosity density ($erg/sec/pc^2$) profile of the galaxy.}
\label{radial}
\end{center}
\end{figure}

\begin{figure}
\begin{center}
\includegraphics[width=3.3in]{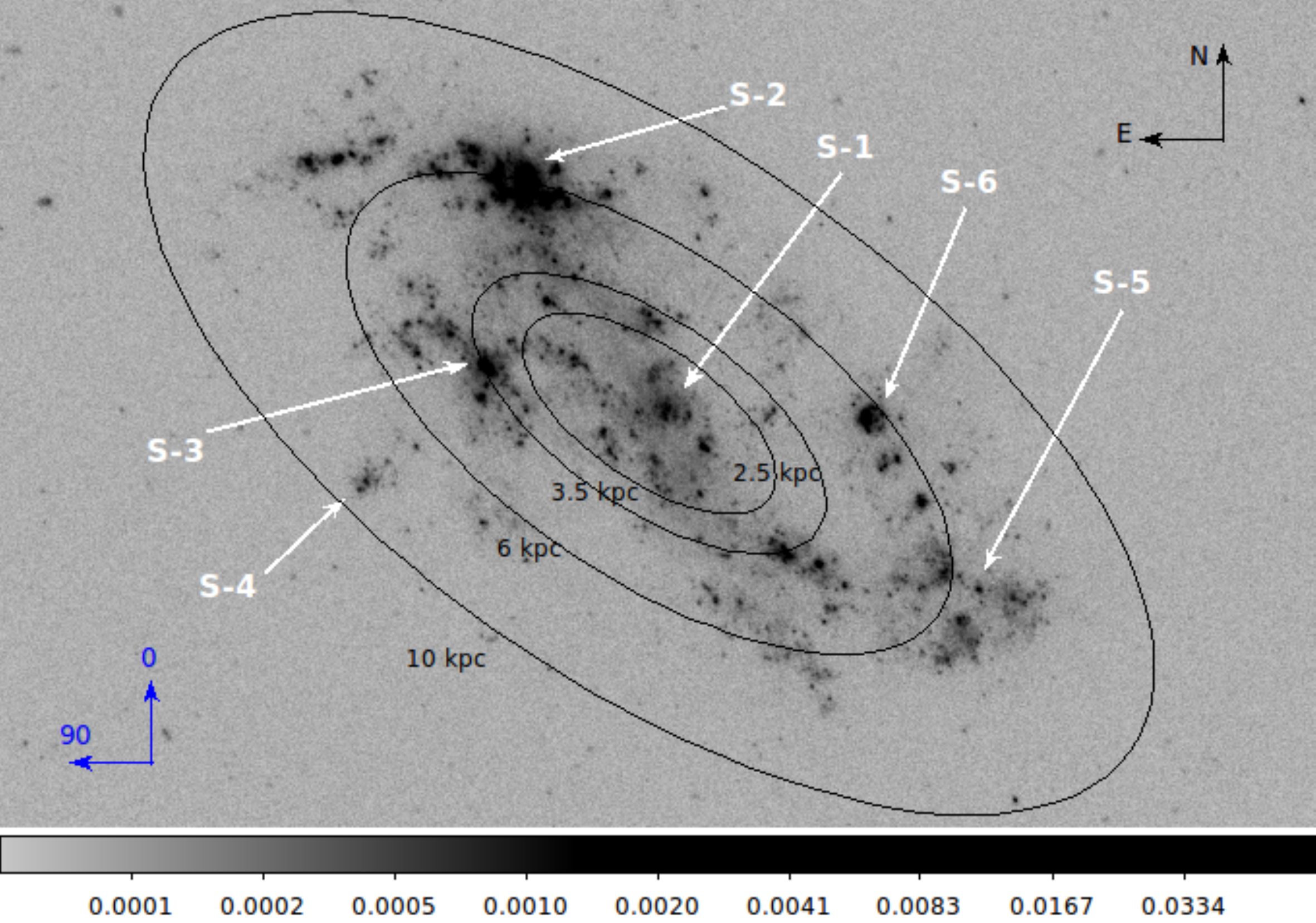} 
\caption{The FUV image of the galaxy IC 2574. The regions highlighted as S1, S2, S3, S4, S5 and S6 are the active star forming regions of the galaxy.}
\label{uv_disk_dist}
\end{center}
\end{figure}

\subsection{FUV Luminosity density profile of IC 2574}
The H~I holes and the bright star forming regions both have a scattered distribution in the galaxy IC 2574. We also noticed the presence of active star formation from inner to far outer part of the galaxy. In order to understand the overall FUV emission profile of the galaxy as a function of galactocentric distance, we plotted radial surface luminosity density ($erg/sec/pc^2$) profile in Figure \ref{radial}. We estimated galactocentric distance to each of the pixel of the FUV image similarly as explained in section \ref{fuv_clump_s}. Starting from the galaxy centre, we considered annulus of width 0.5 kpc up to a radius 10 kpc and estimated total flux in each individual annuli. The measured fluxes are then corrected for background and extinction. The background is estimated from an annuli between radius 12 kpc and 13 kpc whereas extinction correction is done similarly as discussed in Section \ref{ext_s}. The corrected fluxes ($erg/sec/cm^2/\AA$) are converted to luminosity ($erg/sec$) by adopting a distance to the galaxy as 3.79 Mpc \citep{dalcanton2009} and bandwidth of F148W filter as 500 $\AA$ (Table \ref{ic2574} \& \ref{uvit_obs}). Then the estimated luminosity values are divided by the area of each individual annuli to calculate surface luminosity density ($erg/sec/pc^2$) at that particular radius and shown in Figure \ref{radial}. To understand the characteristics of luminosity density profile, we have shown the FUV image of IC 2574 and highlighted the location of six bright star forming regions of the galaxy as S1, S2, S3, S4, S5 and S6 in Figure \ref{uv_disk_dist}. The radial profile shows an exponential decrease up to a radius 2.5 kpc with a central peak which is due to the star forming region S1. Beyond 2.5 kpc, we noticed two peaks, one at 3.5 kpc and another at 6 kpc from the galaxy centre. Emission from the region S3 has major contribution for the peak seen at 3.5 kpc, whereas the peak at 6 kpc is mainly due to S2, the brightest star forming region of the galaxy. The nature of the radial luminosity density profile signifies the presence of active star formation in the outer part of the galaxy also.\\

We also estimated the background and extinction corrected total flux of the galaxy in F148W filter for a radius of 10 kpc and the value is 3.3$\times10^{-12}$ $erg/sec/cm^2/\AA$. The corresponding F148W magnitude and the total SFR (estimated using the relation of \citet{kara2013}) of the galaxy are estimated to be 10.45 mag and 0.57 $M_{\odot}/yr$ respectively. It is to be noted that if we use different relations for estimating SFR from the measured FUV flux, we get different values. For example, the relation provided by \citet{murphy2011} and \citet{hunter2010} results in a SFR of 0.12 and 0.18 $M_{\odot}/yr$ respectively for the same estimated flux of the galaxy IC 2574.\\

\begin{table}
\centering
 
\caption{Starburst99 model parameters}
\label{starburst99}
\resizebox{90mm}{!}{
\begin{tabular}{cc}
\hline

 Parameter & Value\\\hline
 Star formation & Instantaneous\\
 Stellar IMF & Kroupa (1.3, 2.3)\\
 Stellar mass limit & 0.1, 0.5, 120 $M_{\odot}$\\
 Total cluster mass & $10^3 M_{\odot}$-$10^6 M_{\odot}$\\
 Stellar evolution track & Geneva (high mass loss)\\
 Metallicity & Z=0.004\\
 Age & 10 Myr\\ \hline

\end{tabular}
}

\end{table} 
 
\begin{figure}
\begin{center}
\includegraphics[width=3.5in]{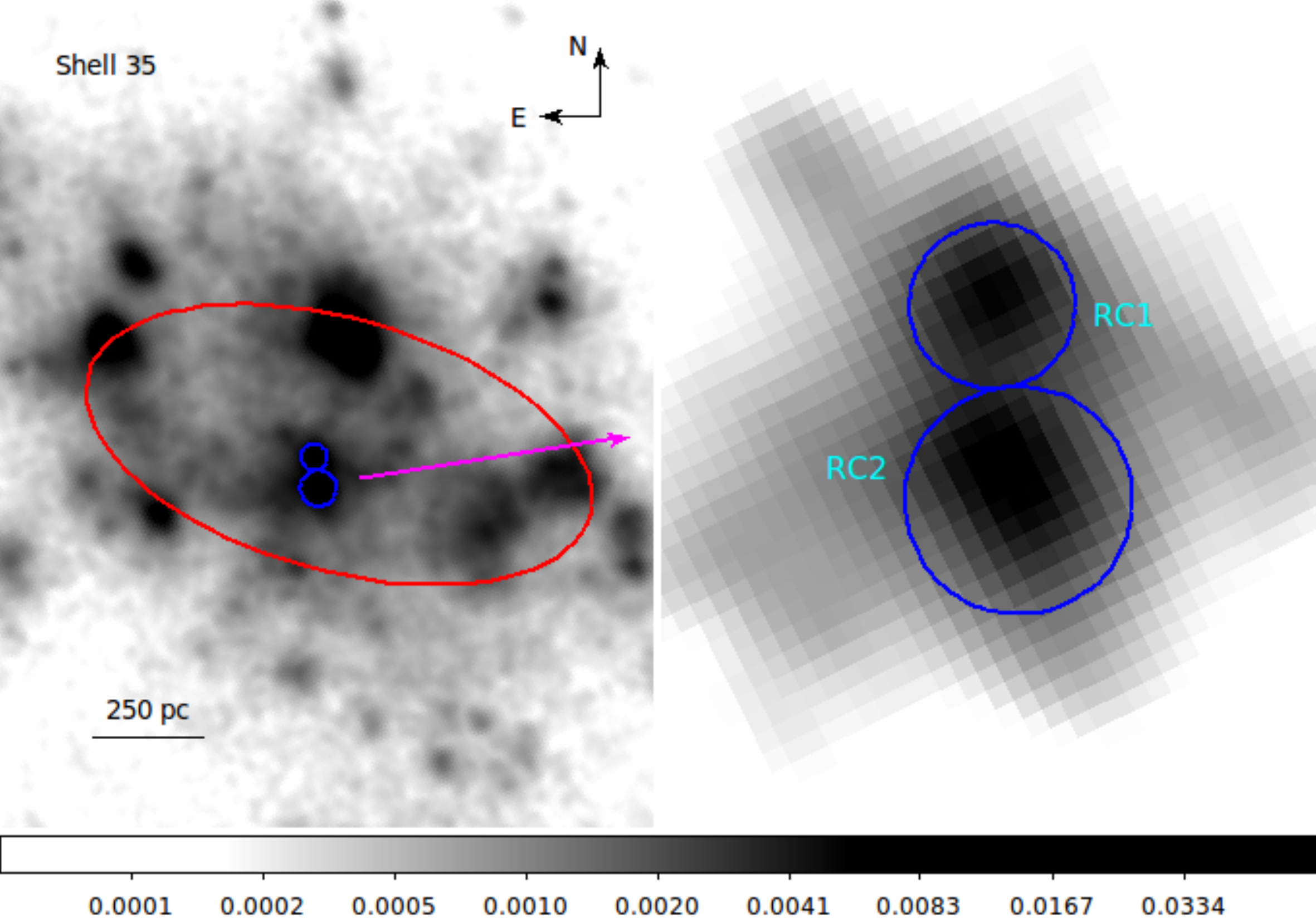} 
 \caption{The FUV image of H$~$I shell 35 (red ellipse) along with the remnant cluster (blue circle) is shown in the left. The zoomed in view of the cluster is shown in right where two resolved components are clearly noticed.}
 \label{sgs35}
 \end{center}
 \end{figure}

\subsection{Remnant cluster of Super Giant Shell}
\label{section_rc}
Among the 48 identified shells, there are multiple SGSs present in the galaxy IC 2574 \citep{walter1999}. One of the present SGSs (shell 35 as per \citet{walter1999}, Figure \ref{shell}), is studied extensively in literature due to its prominent multi-wavelength characteristics. \\

 \begin{table}
\centering
\caption{Properties of the remnant cluster}
\label{cluster_table}
\begin{tabular}{p{1.5cm}p{1cm}p{4cm}p{1.5cm}}
\hline
 Component & Radius  & FUV Flux  & Mass \\
  & (pc) & ($10^{-15}$ $erg/sec/cm^2/\AA$) &($10^4 M_{\odot}$) \\\hline
 RC1 & 30.3 & 7.1 & 4.9\\
 RC2 & 41.7 & 13.2 & 7.3\\\hline
 \end{tabular}
 \end{table}
 
In Figure \ref{sgs35}, we showed the UVIT FUV image of this SGS 35. The presence of remnant cluster as well as star forming regions along the rim are clearly noticed in the image. With the spatial resolution of UVIT, we identified two resolved components of the remnant cluster (RC1 and RC2 in Figure \ref{sgs35}) which were not detected earlier by both UIT and GALEX. The individual components are shown in the figure by blue circular apertures. In order to study the remnant cluster, we estimated the background and extinction corrected FUV fluxes for both the components shown in right panel of Figure \ref{sgs35}. \citet{stewart2000} estimated the age and mass of the central cluster and the reported values are 11 Myr and 14.2$\times10^4$ $M_{\odot}$ respectively. Assuming the age to be 10 Myr, we also estimated the masses for both the components with the help of starburst99 SSP \citep{leitherer1999} model. The model parameters assumed for this estimation are listed in Table \ref{starburst99}. All the measured parameters of the cluster are listed in Table \ref{cluster_table}. The added mass of both the components is $12.2 \times 10^4 M_{\odot}$ which matches closely with earlier estimate by \citet{stewart2000}.

\section{Discussions}
\label{discussion_s}
Giant H~I holes present in some dwarf galaxies represent a prominent feature in their interstellar medium (\citet{egorov2014} and references therein). The slow solid body-like rotation of the dwarf galaxies and the lack of strong spiral density waves, help in the formation of larger sized and long lived holes in these galaxies. Dwarf galaxies are known to sustain star formation over a very long period. It is thus important to understand how the presence of holes and the sustained star formation go hand in hand in dwarf galaxies. The key aim of this study is to identify young star forming regions in the galaxy IC 2574 and further explore their connection with the H$~$I holes. In order to do that we used deep FUV observations of the galaxy with UVIT F148W filter. A 28$\arcmin$ field of view of the telescope along with a pixel scale of 0.4$\arcsec$ has helped us to image the whole galaxy with finer details.\\

 We noticed a good spatial correlation between the FUV emission and the H~I column density throughout the galaxy. The FUV bright regions of the galaxy are mostly found to have H~I column density more than $10^{21} cm^{-2}$,  which is the reported threshold value for star formation in IC 2574. This signifies star formation, in this gas-rich dwarf galaxy, is mainly happening in regions which have higher gas density. We also noticed some active regions to have H~I column density less than  $10^{21} cm^{-2}$. As the FUV emission in some of these regions is high, it is possible that the recent star formation has used up or ionized the neutral gas there. It is also possible that a few of these regions are not part of the galaxy and probably some background sources.\\

We identified 419 star forming regions in the galaxy by fixing the threshold flux as $1.07\times10^{-18}$ $erg/sec/cm^2/\AA{}$ (10 times average background flux) and minimum number of pixel to define a region as 10. A threshold flux lower than the selected value will result in identifying more number of regions and also the same regions with a little bigger size. In case of a higher threshold flux, the identified regions will be less in number and smaller in size. In order to keep a balance, we fixed an intermediate value for our analysis. The value of minimum number of pixel for identifying the regions will decide the minimum size of the region we want to detect. The value we adopted for our analysis can detect star forming clumps smallest up to radius $\sim$ 15 pc. The number of identified regions will increase with decreasing value of the minimum number of pixel and the vice versa. We fixed the value as 10 to identify small star forming clumps and also to avoid detecting very small regions which may not be part of the galaxy.\\

The nature of star formation shows broad characteristic variation for different dwarf galaxies. Both the external environment and internal feedback play dominant role in regulating star formation in dwarfs. Our study highlights that the expanding H$~$I holes have a major impact in triggering secondary star formation in some part of the galaxy. H$~$I holes are mainly created due to the combined effect of stellar winds and supernova explosions on to the ISM of a galaxy \citep{tagle1988}. \citet{tagle2005} concluded that the massive and compact super star clusters contribute a positive feedback for triggering further star formation around them.  We find that star formation could be started either due to scooping of material by an expanding H~I shell and/or collision and compression of matter due to collision of multiple shells. We found that out of 48 holes, 30 show FUV emission in their shells, 15 holes have emission inside and 16 holes do not have any related FUV emission. This denotes that more than half of the holes (30/48) have active star formation in their shells, whereas 16/48 holes do not show any triggered star formation. We also noticed that holes with no related FUV emission mostly lie in the outer part of the galaxy. We found 9 out of 16 holes, which do not show triggered star formation, to be located outside 8 kpc radius of the galaxy, whereas only 1 out of 32 holes, with related FUV emission, is found outside 8 kpc. This signifies that the holes which are formed in the outer part of the galaxy could not trigger star formation. This may happen due to low density of available gas in the far outer part of the galaxy. Our study finds that 28.6\% of the identified star forming regions are located in shells while 12.6\% are present inside. We also found 60.1\% of the regions to be present away from holes and are not related. These numbers are estimated on the basis of our adopted shell width as R/3. The width of the shell can actually be different from our assumed value. This can slightly alter our estimated numbers for each individual holes.\\

In order to identify the star forming regions younger than 10 Myr we checked for their H$\alpha$ counter parts. It turns out that 65 of 120 regions present in the shell show both FUV and H$\alpha$ emission. In other words, 23 holes show both H$\alpha$ and FUV emission in their shells. Therefore, $\sim$ 48\% holes show signature of recently triggered star formation. As per the estimation of \citet{walter1999}, these holes cover an age range of $\sim$ 10 - 40 Myr. If star formation is triggered in some region due the expansion of a hole, then that has to be younger than the age of the hole. As the regions with both FUV and H$\alpha$ emission are likely to be younger than 10 Myr, the detection of these regions in the shell signify that star formation has been triggered there due to the expansion of holes. The regions present in the shells with only FUV emission may be a little older and hence do not show H$\alpha$ emission. Among 252 regions, which are not related with the holes, 106 found to have H$\alpha$ emission also. This means 60.1\% of the identified regions in the galaxy are undergoing star formation triggered due to other mechanisms with 25.3\% experiencing most recent trigger.  Therefore, expansion or collision of H~I holes is not the only mechanism to cause recent enhancement of star formation in the galaxy. The cross-identification of H$\alpha$ emission is done from the available catalog of \citet{miller1994}. The H$\alpha$ observations have $\sim$ 4 times shallower exposure than that in FUV. In the case of some FUV bright regions, there can be faint H$\alpha$ emissions which are not detected in this H$\alpha$ image. Hence, we expect a few more cross-detection with much deeper H$\alpha$ image. That can slightly change the statistics of our results.\\

The radius of the identified regions cover a range between 15 - 285 pc, with 95\% of them to be smaller than 100 pc. The larger regions are the big parent complexes with smaller sub-structures of size 15 - 100 pc inside. As these clumps are bright in FUV, it is possible that some of them are OB associations. The sizes of OB associations cover a range between 10 - 100 pc for the Milky way and other nearby spiral and dwarf galaxies \citep{melnik1995,bresolin1996,ivanov1996,bresolin1998,bastian2007}. Therefore, the galaxy IC 2574 have produced OB associations which are similar in size with those of other nearby galaxies. In order to characterize the star forming regions, we measured their FUV surface luminosity density and the average H$~$I column density. We noticed a clear variation in the properties of these regions across the galaxy. Some regions show very high FUV luminosity with less H$~$I density, signifying vigorous recent star formation. Majority of the regions have moderate FUV luminosity with H$~$I density above the threshold, whereas we also do notice regions with very high H$~$I density and moderate FUV luminosity. These altogether indicate that the galaxy IC 2574 has a variable star forming environment throughout it. In this study, we assumed a fixed extinction value throughout the galaxy and used that to estimate corrected FUV flux, luminosity density and SFR density (Table \ref{uvit_clumps_table}). Any change in the assumed extinction value will affect the estimated values of these parameters accordingly.\\

As it has been suggested that the nature of star forming region is hierarchical from smaller scale to larger scale \citep{elmegreen2000,efremov1995}, we also explored the structural characteristics of star forming regions in the galaxy IC 2574. We noticed that majority of the large star forming complexes in IC 2574 have several smaller sub-structures of different flux levels. It is further observed that the brighter star forming clumps of the galaxy are mostly present inside larger complexes. By analysing the dendrogram of some selected large regions, we understand that in different flux levels, star forming regions form similar structure of different sizes. This highlights the hierarchical nature of these active regions. As the hierarchy is noticed for regions both related and not related to the holes, we speculate turbulence of the ISM as the primary reason behind this \citep{elmegreen2000,elmegreen2014,grasha2017}.\\ 

We produced the radial surface luminosity density profile of the galaxy and able to trace FUV emission at least up to a radius 10 kpc. The presence of two major bright star forming regions, at radius 3.5 kpc and 6 kpc, are identified in this radial profile. The brightest star forming region which is located at 6 kpc from the centre of the galaxy signifies that the star formation in IC 2574 is not concentrated only in the inner part of the galaxy. The presence of this active region clearly reveals that star formation in dwarfs can be dominant in any part of the galaxy depending upon a favourable star forming environment. We estimated the total FUV flux of the galaxy within 10 kpc radius and found it to be 3.3$\times10^{-12}$ $erg/sec/cm^2/\AA$. The measured SFR of the galaxy is 0.57 $M_{\odot}/yr$, which signifies a relatively active nature of IC 2574.\\

With the help of UVIT's spatial resolution, we identified two resolved components of the remnant cluster of shell 35. \citet{stewart2000} reported the discovery of this remnant cluster (single component) using UIT data and derived its properties. We derived the mass for each of the resolved cores by using starburst99 SSP model. The bigger component (M $\sim 7.3 \times 10^4$  $M_{\odot}$) is found to be 1.5 times massive than the smaller component (M $\sim 4.9 \times 10^4$ $M_{\odot}$). The added mass (M $\sim 12.2 \times 10^4$ $M_{\odot}$) of these components matches well with the mass of the remnant cluster measured earlier by \citet{walter2008}.

\section{Summary}
\label{sumarry_s}
The main results of the study are summarized below

\begin{enumerate}
    \item We identified 419 FUV bright star forming regions in the galaxy IC~2574 with the help of UVIT FUV imaging data.
    \item We estimated several parameters, such as size, FUV flux, surface luminosity density, H~I column density, SFR density, galactocentric distance for the identified regions.
    \item We found 28.6\% of the identified regions to be located in H~I shells, 12.6\% inside holes and 60.1\% away from holes.
    \item 30 out of the 48 holes show triggered star formation in their shells with 23 of them having more recent trigger, whereas 16 holes do not show any triggered star formation. We also found 15 holes to have FUV emission inside them, with 12 of those having emission in their shells as well.
    \item Star formation in 60.1\% of the identified regions in IC~2574 has no connection with the H~I holes and hence it is possibly triggered due to other mechanisms.
    \item The identified regions have radii between 15 - 285 pc, with 95\% of them smaller than 100 pc.
    \item 82.3\% of the identified FUV bright regions have H~I column density more than $10^{21} cm^{-2}$.
    \item We found sub-structures of different flux levels and sizes inside the larger star forming complexes across the galaxy. We speculate turbulence as one of the dominant drivers to build this hierarchy.
    \item The galaxy is found to have active star formation in the outer part beyond 5 kpc also.
    \item We resolved two individual components for the remnant cluster of shell 35 and estimated their masses. The added mass of both the component is $\sim 12.2 \times 10^4$ $M_{\odot}$ with the larger one to be 1.5 times more massive than the smaller one.
    \item The star formation rate of the galaxy is found to be $\sim$0.57 $M_{\odot}/yr$.
    
\end{enumerate}

\acknowledgments
UVIT project is a result of collaboration between IIA, Bengaluru, IUCAA, Pune, TIFR, Mumbai, several centres of ISRO, and CSA. Indian Institutions and the Canadian Space Agency have contributed to the work presented in this paper. Several groups from ISAC (ISRO), Bengaluru, and IISU (ISRO), Trivandrum have contributed to the design, fabrication, and testing of the payload. The Mission Group (ISAC) and ISTRAC (ISAC) continue to provide support in making observations with, and reception and initial processing of the data.  We gratefully thank all the individuals involved in the various teams for providing their support to the project from the early stages of the design to launch and observations with it in the orbit. This research made use of Matplotlib \citep{matplotlib2007}, Astropy \citep{astropy2013,astropy2018}, Astrodendro (http://www.dendrograms.org/), community-developed core Python packages for Astronomy. Finally, we thank the referee for valuable suggestions.

\software{CCDLAB \citep{postma2017}, SAOImageDS9 \citep{joye2003}, Matplotlib \citep{matplotlib2007}, Astropy \citep{astropy2013,astropy2018}, Astrodendro (http://www.dendrograms.org/)}


\end{document}